\input{aipcheck}

\documentclass[
    ,final            
  ]
  {aipproc}

\layoutstyle{6x9}

\usepackage{amsmath}
\begin{document}

\title{Vortex reconnections between coreless vortices in binary condensates}



\author{S. Gautam}{
       address = {Indian Institute of Science, Bangalore--560 012, India}
}

\author{K. Suthar}{
  address={Physical Research Laboratory, Ahmedabad--380 009}
}

\author{D. Angom}{
  address={Physical Research Laboratory, Ahmedabad--380 009}
}

\begin{abstract}
   Vortex reconnections plays an important role in the turbulent flows 
   associated with the superfluids. To understand the dynamics, we examine the 
   reconnections of vortex rings in the superfluids of dilute atomic gases 
   confined in trapping potentials using Gross-Petaevskii equation. Further
   more we study the reconnection dynamics of coreless vortex rings, where
   one of the species can act as a tracer.
\end{abstract}
\maketitle


\section{Introduction}
\label{I}
In superfluids, vortex reconnections were first studied theoretically by
Schwarz \cite{schwarz-85,schwarz-88}, who pioneered the vortex 
filament method to study the dynamics of vortices in a superfluid. Later 
Levine et al. \cite{koplik-93} studied the vortex reconnections
using non-linear Schr\"odinger equation as a model for the superfluid system.
Vortex reconnections have been experimentally observed in liquid Helium 
\cite{Bewley}, whereas in Bose-Einstein condensates (BECs) their experimental
realisation is yet to be accomplished. We study the vortex reconnections in 
phase-separated binary condensate of $^{85}$Rb-$^{87}$Rb \cite{papp-08}. In 
this system, the tunability of scattering length of $^{85}$Rb 
\cite{cornish-00} allows us to choose the appropriate interaction strength
suitable for the creation of coreless vortices \cite{gautam-12,gautam-13},
i.e phase-singularities in one of the species with the corresponding density
maxima in the other \cite{catelani-10,mason-11}. Using the phase imprinting
method \cite{} we examine the reconnections between six counter propagating 
coreless vortex rings, and  between a coreless vortex ring and a coreless 
vortex line. For this we consider the system in a spherically symmetric 
trapping potential. It must be mentioned that all the previous studies on 
the vortex reconnections have considered only single species Bose-Einstein 
condensates (BECs). In earlier works, vortex reconnections has been 
studied between counter propagating rings, resulting in the cascade of Kelvin
waves \cite{kivotides-01}. A similar study
\cite{springerlink:10.1007/s10909-010-0287-z} reported the vortex reconnections 
between two rings, and a ring and a line. Further more, the vortex
reconnections between two counter-propagating vortex rings with a slight
offset between their axes and resulting in the emission of sound waves 
was studied in Ref. \cite{leadbeater-01}. Coming to vortex lines, 
reconnections between the bundles of vortex lines in superfluid has been 
studied in Ref. \cite{alamri-08}. All of these studies, however, are with
homogeneous and single species superfluids.  To the best of our knowledge,
the present studies are the first studies on the vortex reconnections 
(between coreless vortices) in binary systems.


\section{Vortex rings in condensates}
In the mean field approximation, a scalar BEC is described by the 
Gross-Pitaevskii (GP) equation
\begin{equation}
 \left[ \frac{-\hbar^2}{2m}\nabla^2 + V({\mathbf r}) + 
 U|\Psi({\mathbf r},t)|^2 - 
 i\hbar\frac{\partial}{\partial t}\right]\Psi ({\mathbf r},t) = 0,
 \label{GP}
\end{equation}
where $U = 4\pi\hbar^2a/m$ with $a$ as the $s$-wave scattering length is 
inter-atomic interaction parameter, and $V({\mathbf r})$ is the trapping 
potential. In the present work, we consider $^{87}$Rb condensate 
\cite{anderson-95} with $10^6$ atoms in a spherical trap 
\begin{equation}
  V(\mathbf r) = \frac{m\omega^2r^2}{2},
\end{equation} 
where $\omega/2\pi = 100$Hz is the trapping frequency. The Eq.~(\ref{GP}) can 
be written in scaled units using 
$a_{\rm osc}= \sqrt{\hbar/m\omega}$, $\omega^{-1}$, and $\hbar\omega$ as the 
units of length, time, and energy respectively. The scaled GP equation is
\begin{equation}
  \left[ -\frac{\tilde{\nabla}^2}{2} + V(\tilde{{\mathbf r}}) + 
  \tilde{U}|\phi(\tilde{{\mathbf r}},\tilde{t})|^2 - 
   i\frac{\partial}{\partial \tilde{t}}\right]
  \phi_i (\tilde{{\mathbf r}},\tilde{t}) = 0,
\label{GP-scaled}
\end{equation}
where $\tilde{\nabla}^2 = a_{\rm osc}^2\nabla^2$, $\tilde{U} = 
4\pi aN/a_{\rm osc}$, and $\phi(\tilde{{\mathbf r}},\tilde{t}) =   
\sqrt{a_{\rm osc}^3/N}\Psi({\mathbf r},t)$ with $N$ as the number of atoms. 
For notational simplicity, we will represent the scaled quantities without 
tilde in the rest of the manuscript. The scaled wavefuntion $\phi$ is 
normalised to unity.

  We first study the reconnections between six vortex rings. The stationary 
state solution is obtained by propagating the coupled Eq.~\ref{GP-scaled} in 
imaginary time. After each time step phase consistent with the presence of 
six vortex rings in $^{87}$Rb is imprinted. Mathematically it implies that 
if $r_0$ is the radius of the six rings located at $x=\pm R_0$, 
$y = \pm R_0$, and $z = \pm R_0$, redefine $\phi(\mathbf r)$ as 
\label{III}
\begin{equation}
\phi = |\phi|   \prod_{i=1}^3\exp\left(i\tan^{-1}\frac{x_i-R_0}
                {\sqrt{\sum_{j=1}^3 x_j^2}-r_0}\right)
                \prod_{i=1}^3\exp\left(-i\tan^{-1}\frac{x_i+R_0}
                {\sqrt{\sum_{j=1}^3 x_j^2}-r_0}\right),
\end{equation}
after each time step. Here $x_1 = x$, $x_2 = y$, and $x_3 = z$ respectively
and $j\ne i$.  The stationary state solution thus obtained is 
shown in Fig. \ref{six_ring}(a).


\subsection{Vortex ring reconnections}

The radius of the core of the vortex ring is approximately equal to the 
coherence length $\xi$. In the point source approximation, we assume the 
vortex ring to be infinitely thin, i.e. $\xi\rightarrow 0$. For such a ring 
with the center located at $(0,0,0)$ and lying on the $xy$ plane in an infinite 
homogeneous medium, the three components of the velocity field are
\begin{eqnarray}
v_x & = & \frac{\Gamma}{4\pi R}
         \int_0^{2\pi} \frac{z \cos \theta}{\left[1+r^2-2(x\cos \theta+y\sin \theta)
          \right]^{3/2}}d\theta,\nonumber\\
v_y & = & \frac{\Gamma}{4\pi R}
         \int_0^{2\pi} \frac{z \sin \theta}{\left[1+r^2-2(x\cos \theta+y\sin \theta)
          \right]^{3/2}}d\theta,\nonumber\\
v_z & = & \frac{\Gamma}{4\pi R}
         \int_0^{2\pi} \frac{1-x\cos \theta-y\sin \theta}{\left[1+r^2-2(x\cos \theta+y\sin \theta)
          \right]^{3/2}}d\theta,\nonumber\\
\label{vel_field}
\end{eqnarray}
where $\Gamma$ is the circulation, $R$ is the radius of the ring, and 
$r=\sqrt{x^2+y^2+z^2}$ is the distance of the observation point from the
center of the ring. Hence, the total velocity field is
\begin{equation}
\mathbf v(\mathbf r) = v_x(\mathbf r)\hat i + v_y(\mathbf r)\hat j 
                       + v_z(\mathbf r)\hat k.
\end{equation}
If there is another ring located at $(t_x,t_y,t_z)$ on $xy$ plane, the total 
velocity field produced by these two rings is
\begin{equation}
\mathbf v(\mathbf r)_t = \mathbf v(\mathbf r) 
                         + \mathbf v(\mathbf r - \mathbf t). 
\end{equation}
In case the ring is located in an arbitrary plane with Euler angles 
$\alpha,\beta,$ and $\gamma$, the velocity field is
\begin{equation}
\mathbf v'(\mathbf r') = R^{-1}(\alpha,\beta,\gamma)\mathbf v(\mathbf r'),
\end{equation}
here rotational matrix 
$R(\alpha,\beta,\gamma) = R_z(\alpha)R_y(\beta)R_z(\gamma)$
and $\mathbf r' = R(\alpha,\beta,\gamma) \mathbf r$. The $R_z(\phi)$ 
and $R_y(\phi)$ define the rotation of the coordinate system about $z$ and 
$y$ axis in three dimensional Euclidean space and are given as 
\begin{equation}
R_z(\phi) = \left( \begin{array}{ccc}
\cos\phi & \sin\phi & 0 \\
-\sin\phi & \cos\phi & 0 \\
0 & 0 & 1 \end{array} \right), \text{ and }
R_y(\phi) = \left( \begin{array}{ccc}
\cos\phi & 0 & -\sin\phi \\
0 & 1 & 0 \\
\sin\phi & 0 & \cos\phi \end{array} \right).
\end{equation}
For a pair of rings with an arbitrary orientation with respect to each other, 
we can always choose the coordinate system in such a way that the origin 
lies at the center of one of the rings and the second ring with its center
at $(t_x,t_y,t_z)$ (say) lies on the plane with Euler angles $\alpha,\beta,$ 
and $\gamma$. The most general expression for the velocity field produced 
by such a pair of rings is
\begin{equation}
\mathbf v(\mathbf r)_t = \mathbf v(\mathbf r)
                         + R^{-1} \mathbf v(R (\mathbf r-\mathbf t)),
\end{equation}
where $R = R(\alpha,\beta,\gamma)$.
In the case of BECs in traps, which are inhomogeneous finite sized system with 
non-zero size of the vortex cores, the velocity field of the condensate can 
be evaluated numerically using the quantum mechanical relation
\begin{equation}
v(\mathbf r, t) = -\frac{\phi^*(\mathbf r,t)\nabla\phi(\mathbf r,t)-
                   \phi(\mathbf r,t)\nabla\phi^*(\mathbf r,t)}{2|
                   \phi(\mathbf r,t)^2|}. 
\end{equation}

\begin{figure}[!ht]
\begin{tabular}{ccc}
\resizebox{45mm}{!}
{\includegraphics[trim = 30mm 5mm 30mm 0mm,clip, angle=0,width=8cm]
                 {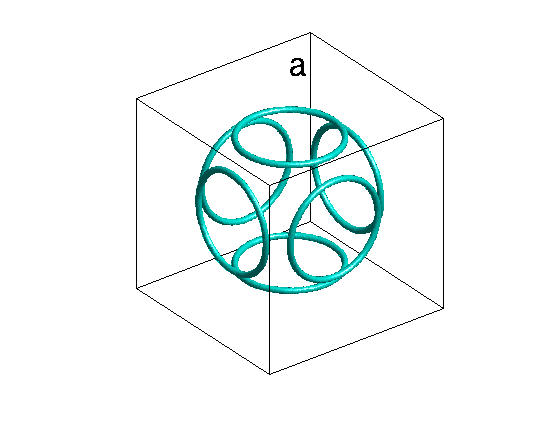}}\!\!\!
\resizebox{45mm}{!}
{\includegraphics[trim = 30mm 5mm 30mm 0mm,clip, angle=0,width=8cm]
                 {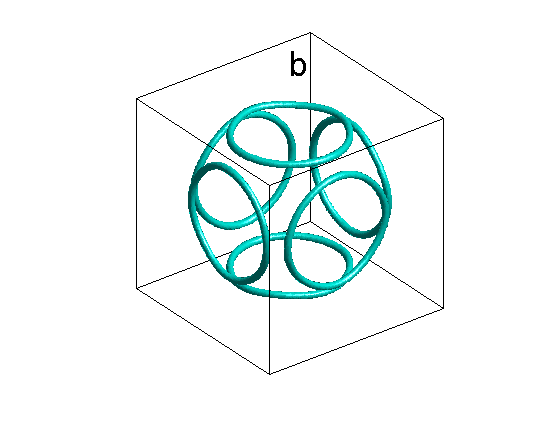}} \!\!\!
\resizebox{45mm}{!}
{\includegraphics[trim = 30mm 5mm 30mm 0mm,clip, angle=0,width=8cm]
                 {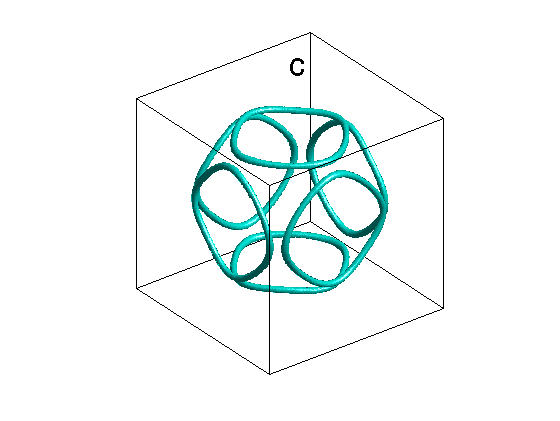}} \\
\resizebox{45mm}{!}
{\includegraphics[trim = 30mm 5mm 30mm 0mm,clip, angle=0,width=8cm]
                 {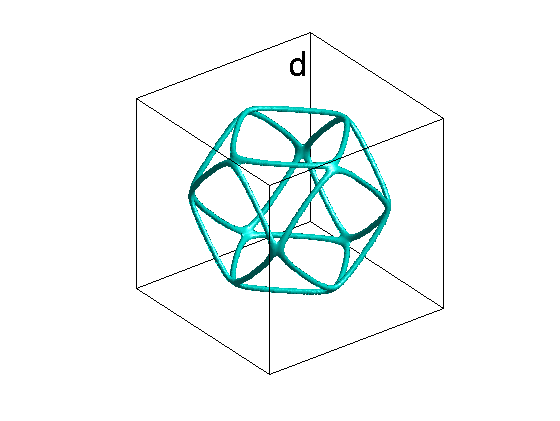}} \!\!\!
\resizebox{45mm}{!}
{\includegraphics[trim = 30mm 5mm 30mm 0mm,clip, angle=0,width=8cm]
                 {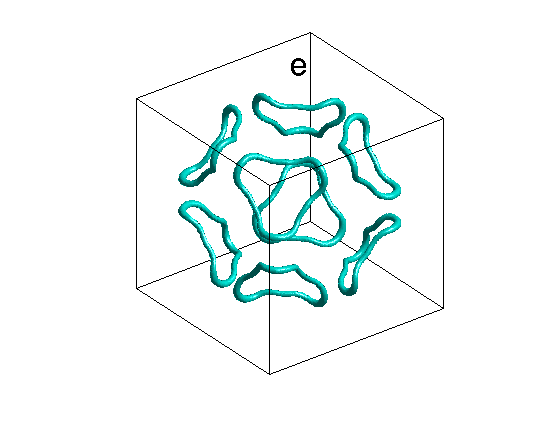}}\!\!\!
\resizebox{45mm}{!}
{\includegraphics[trim = 30mm 5mm 30mm 0mm,clip, angle=0,width=8cm]
                 {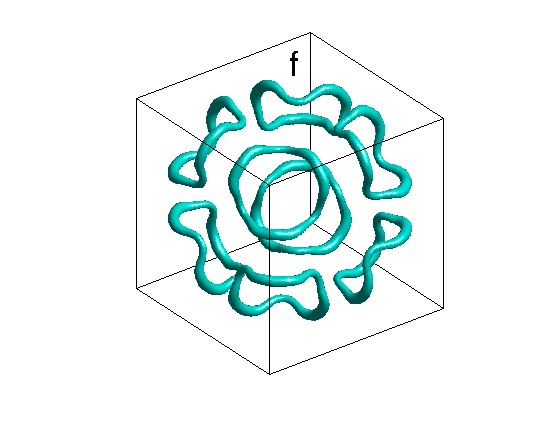}} \\
\end{tabular}
\caption{Reconnections between six vortex rings propagating in opposite
         directions along the three axes.}
\label{six_ring}
\end{figure}

\begin{figure}[ht]
\begin{tabular}{ccc}
\resizebox{30mm}{!}
{\includegraphics[trim = 24mm 0mm 24mm 0mm,clip, angle=0,width=8cm]
                 {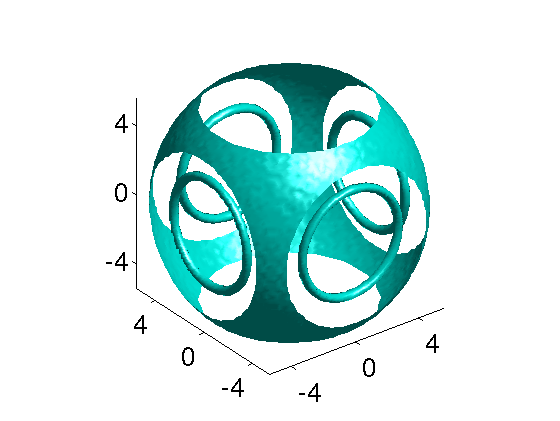}}\!\!
\resizebox{30mm}{!}
{\includegraphics[trim = 24mm 0mm 24mm 0mm,clip, angle=0,width=8cm]
                 {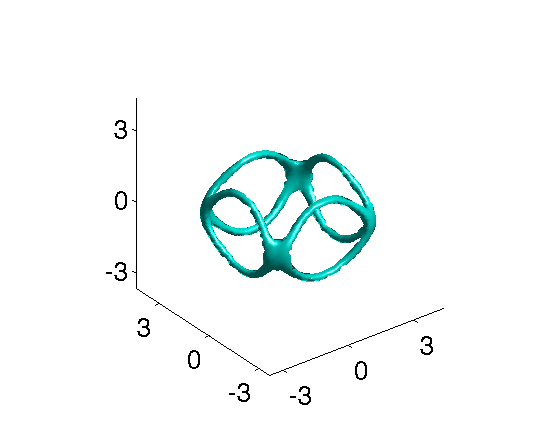}} \!\!
\resizebox{30mm}{!}
{\includegraphics[trim = 24mm 0mm 24mm 0mm,clip, angle=0,width=8cm]
                 {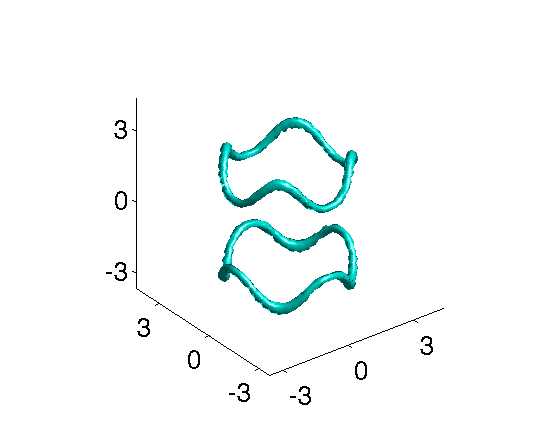}}\\
\resizebox{30mm}{!}
{\includegraphics[trim = 24mm 0mm 24mm 0mm,clip, angle=0,width=8cm]
                 {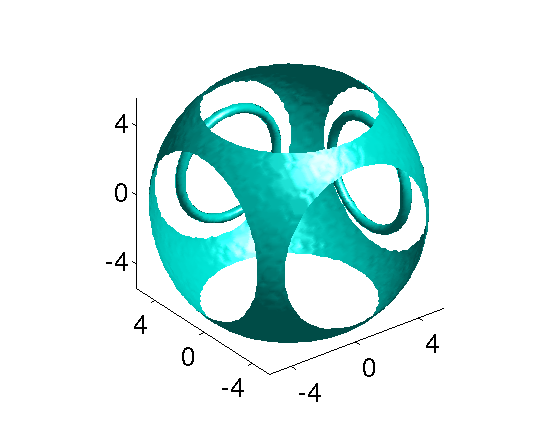}}\!\!
\resizebox{30mm}{!}
{\includegraphics[trim = 14mm 0mm 20mm 0mm,clip, angle=0,width=8cm]
                 {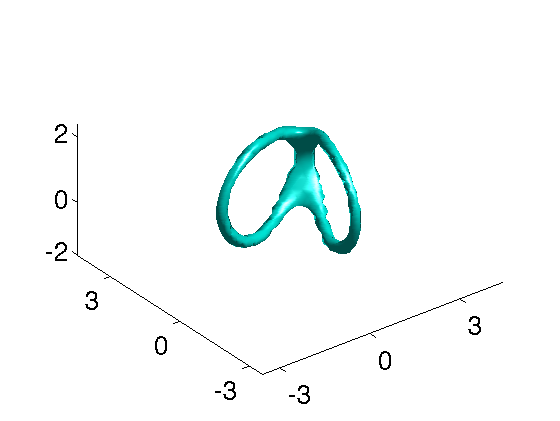}}\!\! 
\resizebox{30mm}{!}
{\includegraphics[trim = 24mm 0mm 24mm 0mm,clip, angle=0,width=8cm]
                 {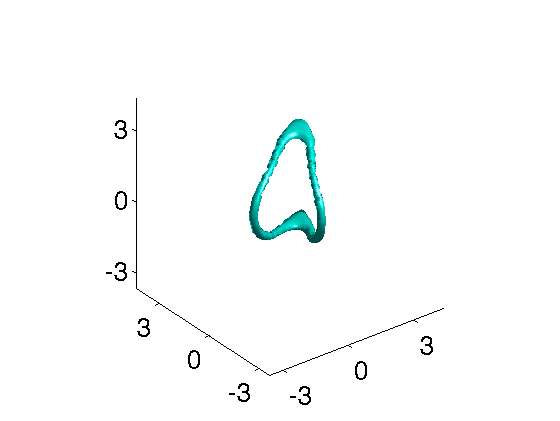}}\\
\end{tabular}
\caption{First reconnection event for four counter-propagating rings
(upper row) and two orthogonally moving rings (lower row).} 
\label{two_four_rings}
\end{figure}

 The velocity field can be extended to the case of six vortex rings approaching
each other along the three axes. Due to the mutual force between the 
neighbouring vortices, the portions of the  rings along the orthogonal 
directions are distorted around the points of least separations. This 
stretches the vortex rings to a rectangular shape and this is clearly 
discernible in Fig. \ref{six_ring}(b-c). The reconnection events sets in when
the rings comes in contact (in Fig. \ref{six_ring}(d) ) and separates into
eight smaller vortex rings after the reconnections, these are shown in 
Fig. \ref{six_ring}(e-f). After the reconnections, the smaller rings propagate 
towards the periphery of the condensate and gets reflected back.
To examine the nature of the reconnection events, we have also studied the 
vortex reconnections between four and two rings. The isosurfaces of $^{87}$Rb 
with $ |\phi_2(x,y,z)|= 0.015$ after the first reconnection event in these 
two cases are shown in Fig.~\ref{two_four_rings}. Topologically, there is a
significant difference in the reconnection dynamics compared to the case of 
six vortex rings, however, there are several common qualitative features. One
of the important pre-reconnection development is the stretching and distortion
of the vortex rings. Regardless of the number of vortex rings, this is 
observed in all the reconnections, 


\section{Phase-separated binary condensate in spherical traps}
\label{II}
As has been mentioned in the {\em introduction}, we consider phase-separated 
binary condensate of $^{85}$Rb-$^{87}$Rb in the spherically symmetric traps. 
The dynamics of this binary system at absolute zero can be studied by using 
coupled Gross-Pitaevskii (GP) equations 
\begin{equation}
 \left[ \frac{-\hbar^2}{2m}\nabla^2 + V_i({\mathbf r}) + 
 \sum_{j=1}^2U_{ij}|\Psi_j({\mathbf r},t)|^2 - 
 i\hbar\frac{\partial}{\partial t}\right]\Psi_i ({\mathbf r},t) = 0,
 \label{gp}
\end{equation}
where $i=1$ for $^{85}$Rb and $i=2$ for $^{87}$Rb. Here
$U_{ii} = 4\pi\hbar^2a_{ii}/m_i$, where $m_i$ is the mass and $a_{ii}$ is
the $s$-wave scattering length, is the intra-species interaction,
$U_{ij}=2\pi\hbar^2a_{ij}/m_{ij}$, where $m_{ij}=m_i m_j/(m_i+m_j)$ is the
reduced mass and $a_{ij}$ is the inter-species $s$-wave scattering length, 
is the inter-species interaction, and $V_i({\mathbf r})$ is the trapping 
potential for the $i$th species. Without loss of generality, we consider equal 
trapping potential for the two species, i.e.
\begin{equation}
 V_i({\mathbf r}) = \frac{m\omega^2}{2}(x^2 + y^2 + z^2),
\label{potentials}
\end{equation}
where $m=m_1\approx m_2$. In the present work, we consider 
$\omega = 2\pi 100.0$Hz as the trapping frequency, $a_{11} = 450a_0$ 
\cite{cornish-00},  $a_{22} = 99a_0$ \cite{kempen-02}, and
$a_{12} = 214a_0$ \cite{burke-99} as the scattering length values and
$N_1 = 5\times10^5$ and $N_2=8\times10^5$ as the number of atoms. With these
parameters, the ground state of the binary system is phase-separated with 
$^{85}$Rb forming a shell around $^{87}$Rb. The Eq.~(\ref{gp}) can be written
in scaled units using $a_{\rm osc}=\sqrt{\hbar/m\omega}$, $\omega^{-1}$, and
$\hbar\omega$ as the units of length, time, and energy respectively. The scaled
GP equation is
\begin{equation}
  \left[ -\frac{\tilde{\nabla}^2}{2} + V(\tilde{{\mathbf r}}) + 
  \sum_{j=1}^2\tilde{U}_{ij}|\phi_j(\tilde{{\mathbf r}},\tilde{t})|^2 - 
   i\frac{\partial}{\partial \tilde{t}}\right]
  \phi_i (\tilde{{\mathbf r}},\tilde{t}) = 0,
\label{gp_scaled}
\end{equation}
where $\tilde{\nabla}^2 = a_{\rm osc}^2\nabla^2$, 
$\tilde{U}_{ii} = 4\pi a_{ii}N_i/a_{\rm osc}$,
$\tilde{U}_{ij} = 4\pi a_{ij}N_j/a_{\rm osc}$, and 
$\phi_{i}(\tilde{{\mathbf r}},\tilde{t}) =   
\sqrt{a_{\rm osc}^3/N_i}\Psi_i({\mathbf r},t)$. For notational simplicity, we
will represent the scaled quantities without tilde in the rest of the 
manuscript.


\section{Coreless vortex ring reconnections}
\label{III}
\begin{figure}[ht]
\begin{tabular}{cc}
\resizebox{45mm}{!}
{\includegraphics[trim = 30mm 5mm 30mm 0mm,clip, angle=0,width=8cm]
                 {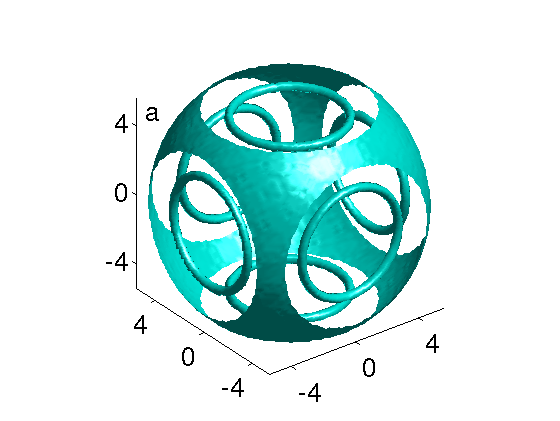}}\!\!
\resizebox{45mm}{!}
{\includegraphics[trim = 30mm 5mm 30mm 0mm,clip, angle=0,width=8cm]
                 {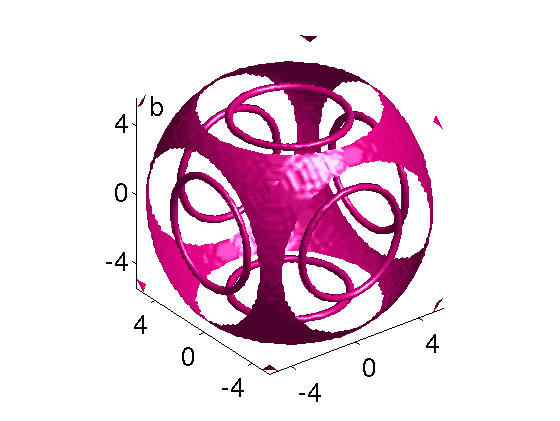}} \\
\end{tabular}
\caption{Isosurfaces of $^{87}$Rb (left image) and $^{85}$Rb (right image)
with $ |\phi_i(x,y,z)|= 0.015$. There are six coreless vortex rings in $^{87}$Rb.
}
\label{figure-1}
\end{figure}

 In binary condensates, the coreless vortices \cite{gautam-12,gautam-13} occur
when the core of the vortices associated with one of the species is occupied 
by the other species. This is reminiscent of using Hydrogen as tracer of 
vortices in superfluid Helium \cite{Bewley} to observe vortex dynamics 
including reconnection events.  To study the reconnections between coreless 
vortex rings, the stationary state solution is obtained by propagating the 
coupled Eqs.~\ref{gp_scaled} in imaginary time. After each time step phase 
consistent with the presence of six vortex rings in $^{87}$Rb is imprinted. 
Mathematically it implies that if $r_0$ is the radius of the six rings located 
at $x=\pm R_0$, $y = \pm R_0$, and $z = \pm R_0$, redefine $\phi_2(\mathbf r)$ 
as 
\begin{equation}
\phi_2 = |\phi_2|   \prod_{i=1}^3\exp\left(i\tan^{-1}\frac{x_i-R_0}
         {\sqrt{\sum_{j=1}^3 x_j^2}-r_0}\right)
         \prod_{i=1}^3\exp\left(-i\tan^{-1}\frac{x_i+R_0}
         {\sqrt{\sum_{j=1}^3 x_j^2}-r_0}\right), 
\end{equation}
after each time step. Here $x_1 = x$, $x_2 = y$, and $x_3 = z$ respectively 
and $j\ne i$. The stationary state solution thus obtained is shown in 
Fig.\ref{figure-1}. 

\begin{figure}[ht]
\begin{tabular}{cc}
\resizebox{45mm}{!}
{\includegraphics[trim = 30mm 5mm 30mm 0mm,clip, angle=0,width=8cm]
                 {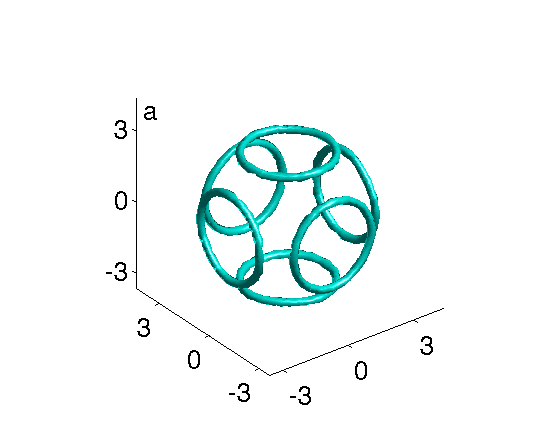}}\!\!
\resizebox{45mm}{!}
{\includegraphics[trim = 30mm 5mm 30mm 0mm,clip, angle=0,width=8cm]
                 {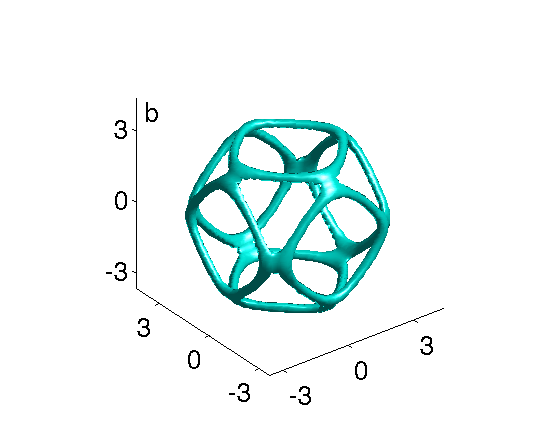}} \\
\resizebox{45mm}{!}
{\includegraphics[trim = 30mm 5mm 30mm 0mm,clip, angle=0,width=8cm]
                 {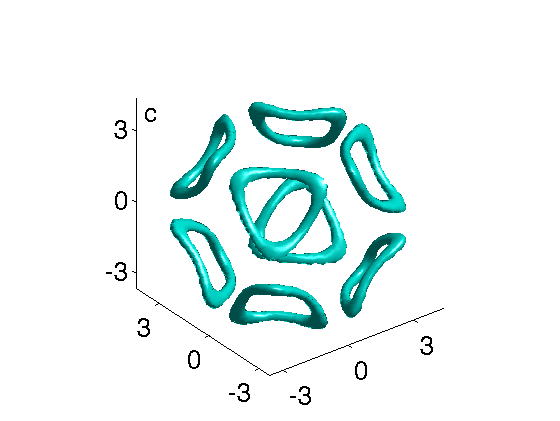}}\!\!
\resizebox{45mm}{!}
{\includegraphics[trim = 30mm 5mm 30mm 0mm,clip, angle=0,width=8cm]
                 {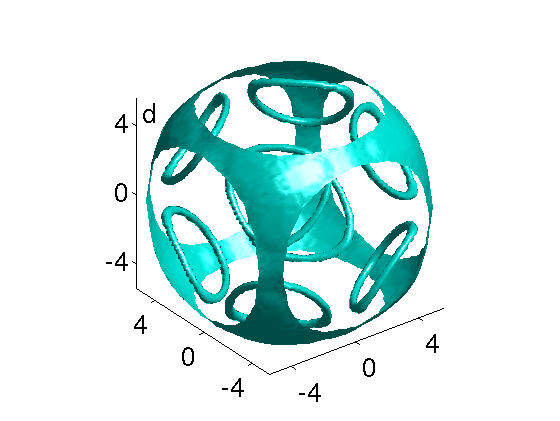}} \\
\end{tabular}
\caption{Isosurfaces of $^{87}$Rb 
with $ |\phi_2(x,y,z)|= 0.015$. There are six coreless vortex rings in $^{87}$Rb
which undergo vortex reconnections leading to formation of eight vortex rings. 
In upper row, the left image is at $8$ ms and the right one is at $9$ ms.
The stretching of the vortex rings is evident from the images. In the
lower row, the left image is at $10$ ms, which shows the presence of Kelvin
waves generated after reconnections, and the right one is at $15$ ms.
}
\label{figure-2}
\end{figure}

 As the vortex rings approach each other, there are four distinct reconnection
events. First, the six vortex rings start moving towards the
center and undergo vortex reconnection at about $11$ ms. Just prior to 
reconnections there is the stretching of vortex rings as is evidenced by the 
visual comparison of two images in the upper row of Fig. \ref{figure-2}. This 
is consistent with previous theoretical studies. Like in the previous case
of reconnections in single species condensate, the reconnections leads to the 
formation of eight daughter vortex rings.

\begin{figure}[ht]
\begin{tabular}{ccc}
\resizebox{30mm}{!}
{\includegraphics[trim = 24mm 0mm 24mm 0mm,clip, angle=0,width=8cm]
                 {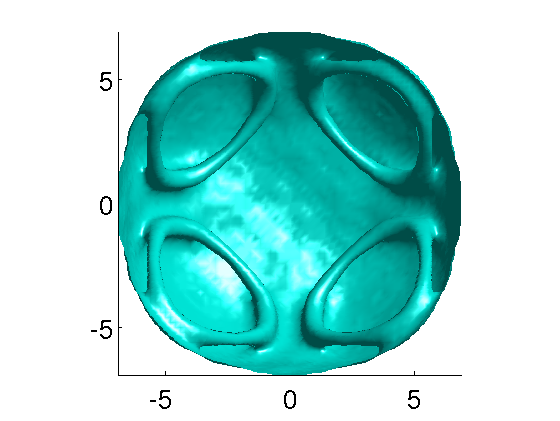}}\!\!
\resizebox{30mm}{!}
{\includegraphics[trim = 24mm 0mm 24mm 0mm,clip, angle=0,width=8cm]
                 {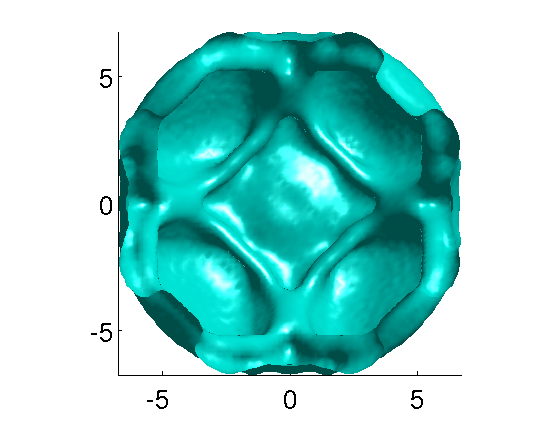}} \!\!
\resizebox{30mm}{!}
{\includegraphics[trim = 24mm 0mm 24mm 0mm,clip, angle=0,width=8cm]
                 {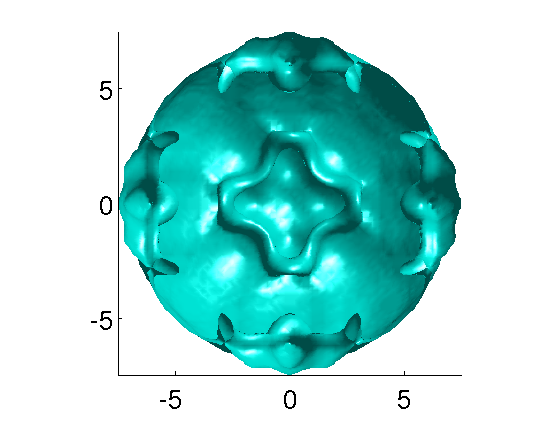}}\\
\resizebox{30mm}{!}
{\includegraphics[trim = 24mm 0mm 24mm 0mm,clip, angle=0,width=8cm]
                 {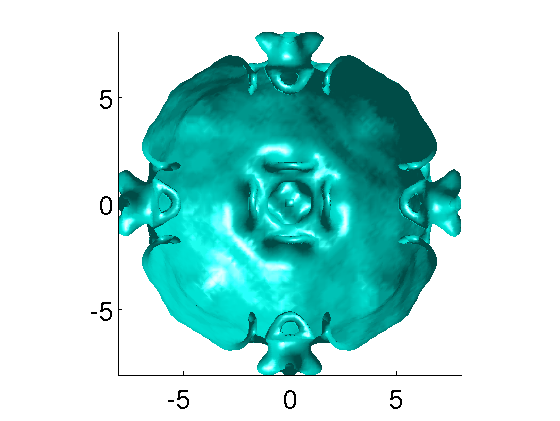}} \!\!
\resizebox{30mm}{!}
{\includegraphics[trim = 24mm 0mm 24mm 0mm,clip, angle=0,width=8cm]
                 {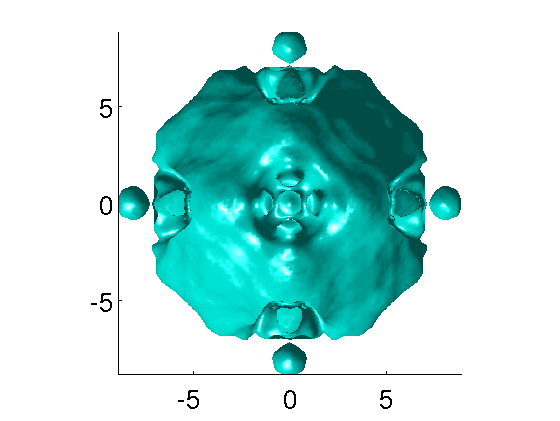}}\!\!
\resizebox{30mm}{!}
{\includegraphics[trim = 24mm 0mm 24mm 0mm,clip, angle=0,width=8cm]
                 {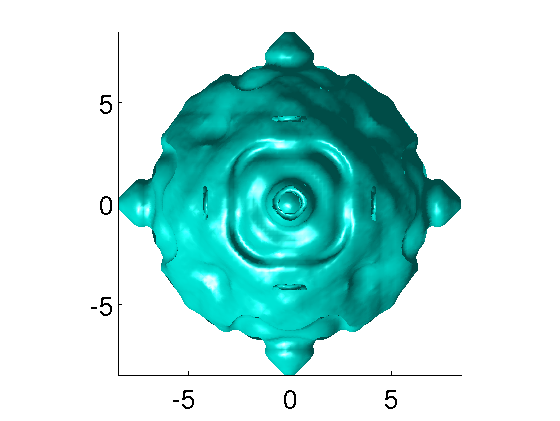}} \\
\end{tabular}
\caption{Half of the isosurfaces of $^{87}$Rb 
with $ |\phi_2(x,y,z)|= 0.015$. Middle images in upper and lower row show
the second and third reconnections events taking place at $17$ ms and $20$ ms
respectively.}
\label{figure-3}
\end{figure}

In the second and third reconnection event, the eight daughter vortex rings 
formed after the first reconnection event move towards the interface, i.e. 
away from the the center of the trap. As they move away, there size increases.
On the interface, the region where the two species overlap, there is a second 
reconnection event as is shown in the middle image in the upper row of 
Fig.~\ref{figure-3}. Again the resultant vortices undergo reconnections to 
form six small vortex rings; this is shown
in the lower row of Fig.~\ref{figure-3}, where one complete ring and halves 
of the four other rings are shown in the rightmost image in the bottom row. 

Finally, in the fourth reconnection event,  the four small rings again start 
moving towards the origin and undergo reconnections after $23$ ms. Like in the
after first reconnection event, there is again the formation of eight smaller 
vortex rings which again start moving towards the interface.
\begin{figure}[ht]
\begin{tabular}{cc}
\resizebox{45mm}{!}
{\includegraphics[trim = 30mm 5mm 30mm 0mm,clip, angle=0,width=8cm]
                 {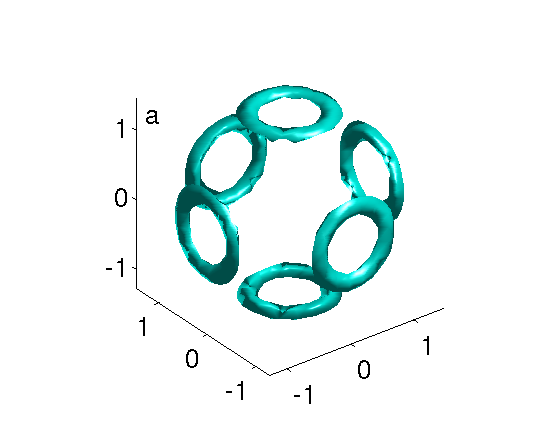}}\!\!
\resizebox{45mm}{!}
{\includegraphics[trim = 30mm 5mm 30mm 0mm,clip, angle=0,width=8cm]
                 {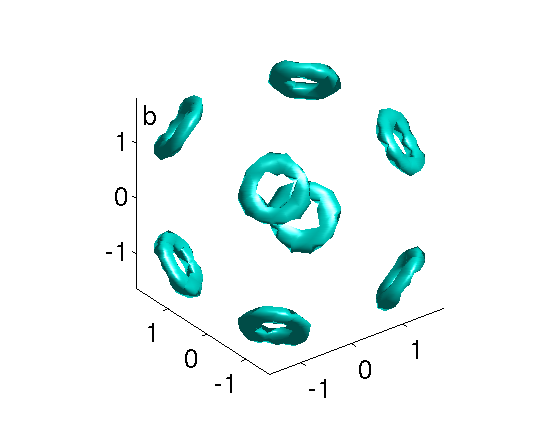}} \\
\end{tabular}
\caption{Isosurfaces of $^{87}$Rb 
with $ |\phi_2(x,y,z)|= 0.015$ at $23$ ms (left image) and $24$ ms (right image). 
}
\label{figure-4}
\end{figure}
It is to be noted that the dynamics of the vortex reconnections in superfluids
show one qualitative difference from the situation in viscous fluids. In the 
latter, the reconnection occurs through {\em bridging}, where the vortices 
after reconnection and separation retain a thin link \cite{kida-94} but this
is absent in superfluids. Another point is, a dissipative process is 
prerequisite for vortex reconnections to occur in superfluids. So, in the 
BECs of trapped atomic gases we need to examine the origin and mechanism of the 
dissipative process. In BEC with inhomogeneous confining
potentials, like in the present study, a vortex precess along equipotential
surfaces. This leads to acceleration of the vortex elements and dissipates
energy through acoustic radiation \cite{kambe-86}. In the case of vortex 
rings, in addition to the self induced velocity of the ring, there is an 
acceleration arising from the inhomogeneity in the confining potential and 
this causes acoustic radiation. So, in superfluids the acoustic radiation is 
an important dynamical mechanism of energy dissipation and can play a 
fundamental role in vortex reconnections. In fact, one of our ongoing 
study is to examine this process in more detail. The other dynamical effect 
which could contribute to the dissipation is the interaction of the vortices 
with the background excitations. This, however, is likely to be less 
important for the present work as the studies are at zero temperature. In 
conclusion, we have demonstrated that coreless vortex rings in binary 
condensates undergo vortex reconnections and in spherical geometry, there are 
multiple episodes of reconnections.


\begin{theacknowledgments}
We thank S. Chattopadhyay and Arko Roy for very useful 
discussions. The numerical computations reported in the paper were done on
the 3 TFLOPs cluster at PRL. 
\end{theacknowledgments}

\bibliography{vortex_connections} 
\end{document}